\documentclass{webofc}

\usepackage[varg]{txfonts}   
\begin{document}
\title{Development of a jet gas target system for the Felsenkeller underground accelerator}

\author{\firstname{Anup} \lastname{Yadav}\inst{1,2}\fnsep\thanks{\email{a.yadav@hzdr.de}}
        \firstname{Konrad} \lastname{Schmidt}\inst{1}\fnsep\thanks{\email{konrad.schmidt@hzdr.de}} 
        \firstname{Daniel} \lastname{Bemmerer}\inst{1,2}
}

\institute{Helmholz-Zentrum Dresden-Rossendorf, Institute of Radiation Physics, 01328 Dresden, Germany
\and
           Technische Universit{\"a}t Dresden, 01069 Dresden, Germany 
          }

\abstract{
 For direct cross section measurements in nuclear astrophysics, in addition to suitable ion beams and detectors, also highly pure and stable targets are needed. Here, using a gas jet as a target offers an attractive approach that combines high stability even under significant beam load with excellent purity and high localisation. Such a target is currently under construction at the Felsenkeller underground ion accelerator lab for nuclear astrophysics in Dresden, Germany. The target thickness will be measured by optical interferometry, allowing an in-situ thickness determination including also beam-induced effects. The contribution reports on the status of this new system and outlines possible applications in nuclear astrophysics.
}

\maketitle
\section{Introduction}
The Felsenkeller underground laboratory is suitable to study, amongst others, two important astrophysical reactions of interest: ${^3}$He($\alpha,\gamma$)${^7}$Be and ${^{12}}$C($\alpha,\gamma$)${^{16}}$O. The first reaction controls the hydrogen burning in the Sun and also affects lithium production in Big Bang Nucleosynthesis (BBN). In order to measure the astrophysical S-factor and angular distribution of this reaction, we are planning to use a ${^3}$He gas jet target for comprehensive data coverage of the entire BBN range. 

The ${^{12}}$C($\alpha,\gamma$)${^{16}}$O reaction affects the abundance ratio of the two elements carbon and oxygen. Due to the low cross-section in the Gamow window, significant statistics for this reaction will be achieved with ${^{12}}$C beam incident on a static windowless ${^4}$He gas target, and a 4$\pi$ detector.

To measure reactions of astrophysical interest targets are needed that are locally well confined, highly dense, thermally stable, and chemically pure. 
Measurements of the angular distribution require a highly localized target for a high-resolution of the reaction energy and the emission angle.
A solid target often contains other elements through contaminations and requires backing materials like tantalum for mechanical stability, which may increase the beam induced background and energy straggling in case of implantation. Solid targets can also degrade over time or exposure to a beam resulting in issues like overheating, and blistering.

A wall-jet gas-target is an alternative solution, which can provide a high gas density, high localization, chemically purity, and a thermal stability \cite{KSchmidt,2,4,5}.
\begin{figure}[htbp]
\centering
	\includegraphics[width=8cm,height=6cm]{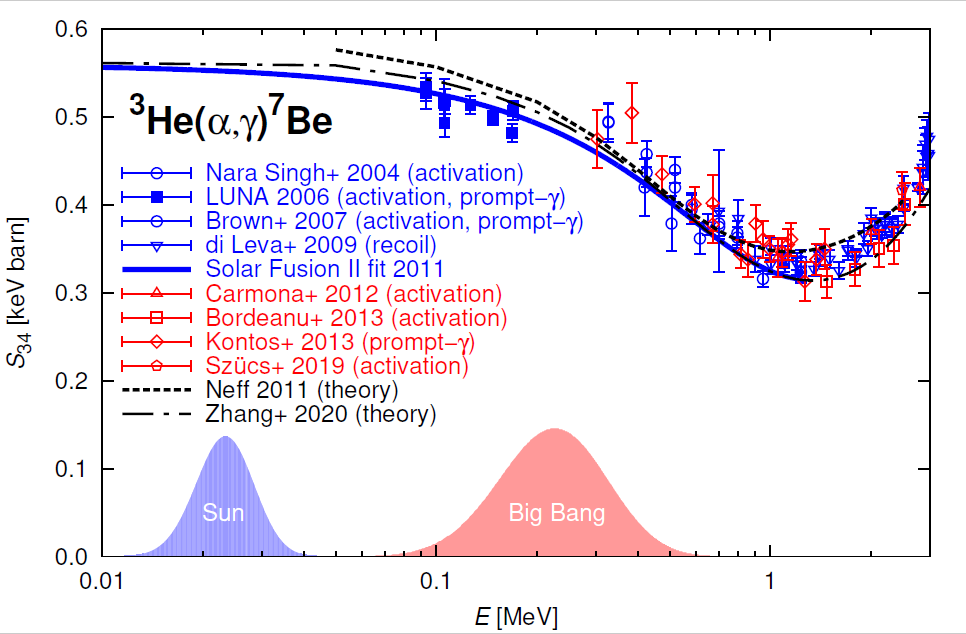} 
	\caption{Astrophysical S-factor of the ${^3}$He($\alpha,\gamma$)${^7}$Be reaction as a function of energy. The Felsenkeller gas-jet will help to perform comprehensive measurements of the reaction covering the entire BBN range.}
	\label{fig:1}
\end{figure}

\section{Experimental setup}

\subsection{Differential pumping system}
\begin{figure}[htbp]
\centering
	\includegraphics[width=10cm,height=5cm]{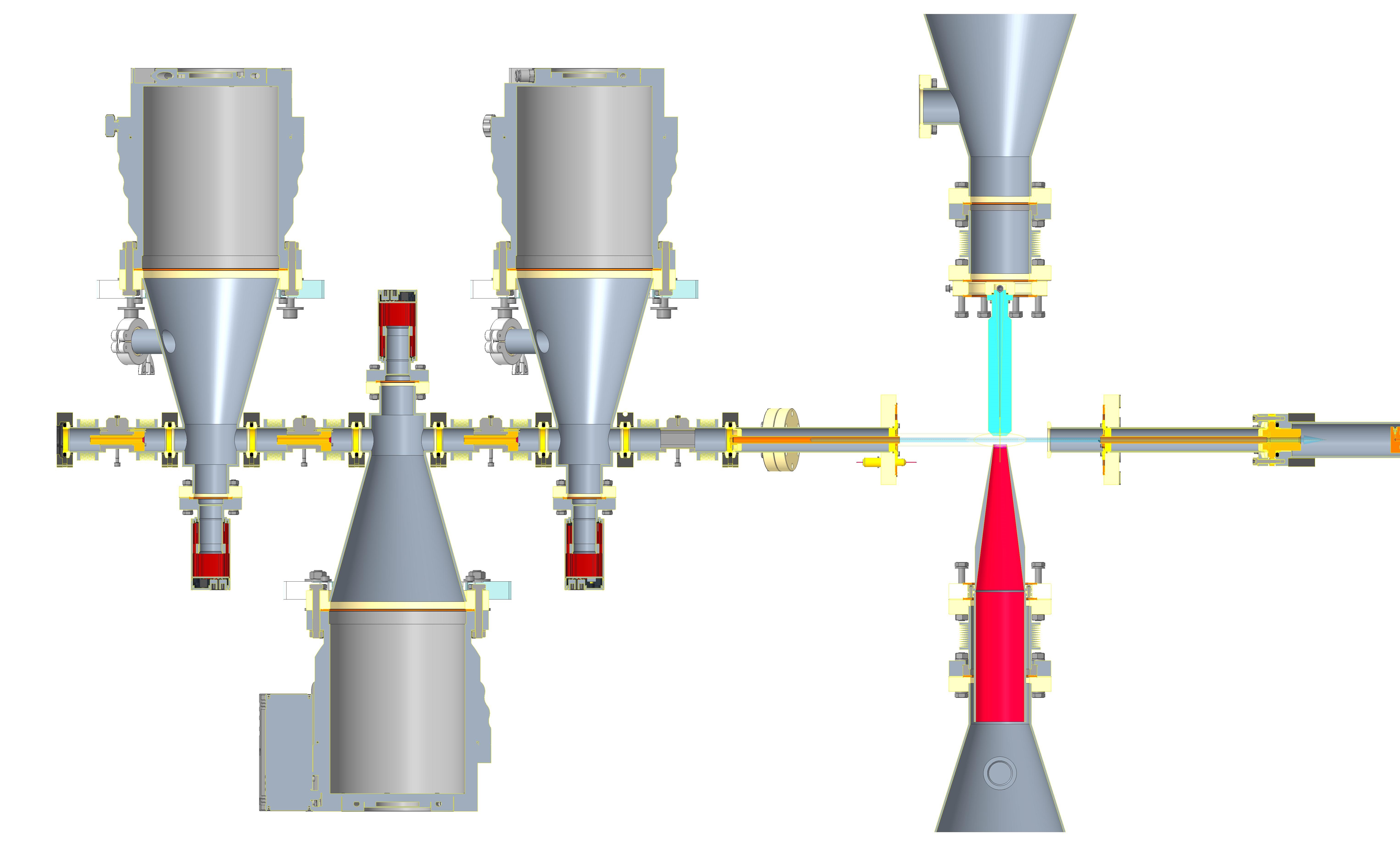} 
	\caption{Technical drawing of differential pumping stages. The beam enters at the left and passes through apertures in between the pumping stages, the jet chamber and the static on the most right where it is stopped in a beam calorimeter. The pumping stages are evacuated by three 700 l/s and the jet chamber by a 2300 l/s turbomolecular pump.}
	\label{fig:2}
\end{figure}
A recirculating gas-jet target (Fig. 2) creates a supersonic jet based on the JENSA gas jet target \cite{KSchmidt}. Here, the target density is 10 times less than at JENSA due to the different physics case. The exhaust of the nozzle has a cross-sectional area at the neck of less than 1 mm. About 12 mm below the nozzle a cone-shaped gas catcher receives most of the gas jet. Catcher is connected to a cone-shaped chamber, and this chamber is evacuated by a 2300 liter turbomolecular pump. The exhaust of all the pumps are connected to a multi-stage root pump and this pump will then compress the gas back to the initial inlet pressure. Before return back to the nozzle, the gas is also passing a chemical getter that will clean it from residual contamination that might have enter during the circulation.

Part of the gas jet diffusing in the jet chamber and return to the beam line. This residual gas is evacuated by a 2300 liter turbomolecular pump on the top of the jet chamber and three 700 liter turbomolecular pumps along the beam line, all separated by copper apertures. These collimators reduce the pressure down to $10^{-7}$ mbar, that is necessary for the ion accelerator. Further, the apertures also reduce the beam size to a diameter of 5 mm or less, in order to provide a well defined beam spot. Downstream of the jet, a extended static gas chamber with a beam calorimeter (see the section 2.3) is installed to determine the ion beam intensity.

In order to monitor the system, all the pumps, pressure gauges, and valves are controlled by a LabVIEW system and parameters of the system are logged. The gas system can be operated either in jet mode or in static mode.

\subsection{Wall jet gas target}
For measurements of low cross-section reactions, a high density of target nucleui within the confined region is needed to maximize count rate. On the other hand, a limited target thickness reduces reaction product energy loss and straggling. The size of the jet should be comparable to the full beam spot so the beam crosses the jet within the homogeneous center. The inlet pressure should be sufficiently high to get a supersonic jet. The characteristics of the gas wall jet that are aimed for, are summarized in Table \ref{tab-1}.

The nozzle is placed vertically with the jet flowing downward with two inlet connections. The shape of the nozzle is a converging-diverging, de Laval type of nozzle [6]. The distance between the nozzle and catcher is around 12 mm as suggested by the literature \cite{4, 8} and by the beam size. This value can be changed for future measurements.

\begin{table}
\centering
\caption{Aimed characteristics of the gas wall jet}
\label{tab-1}       
\begin{tabular}{lll}
\hline
Parameter & Aim & Remark\\
\hline
Areal density & $ 10^{18}$ atoms/$\mathrm{cm}^{2}$ & Single resonance measurement \\
Thickness & <1 mm & Reaction localization for angular distribution measurements\\
Width &  10 mm & Beam (5 mm diameter) crosses jet within homogeneous center \\
Inlet pressure & 	$\sim$1 bar & Needed to get a supersonic jet \\
Mass flow & 1.4 bar l/s & Result of bottle neck cross section of $\sim$1 mm${^2}$  \\\hline
\end{tabular}
\end{table}

\subsection{Static gas target with beam calorimeter}
Downstream of the gas wall jet, a windowless, extended gas target of static type (Fig. 3) is installed that contains the beam calorimeter. The concept is based on the LUNA gas target[2]. In static mode, the gas enters the target chamber from the right through the pipes shown in grey color. The gas flow is regulated by two  parallel metered valves (EVR 116) to keep a constant pressure in the chamber. The expected target pressure is around 2.0 mbar.

A collimator between the jet target chamber and the entrance of extended target limits the gas flow in the direction of the jet chamber. Any remaining target gas is removed by the jet chamber pumps and three differential pumping stages (see section 2.1).

It is not possible to use a Faraday cup to measure the beam intensity while the target chamber is filled with gas because of secondary electrons, therefore a different technique in form of a power compensation calorimeter is used. The calorimeter is made from copper and consists of hot side with heating resistor (12$\Omega$, 60W) inside the copper structure and cold side. The hot and cold sides are always kept at a constant temperature with the help of temperature regulated heating resistors at the hot side and circulating low-temperature coolant from a high precision temperature stability chiller (Julabo 600F) at the cold side. When the ion beam hits the calorimeter, the resistor will reduce the heating to maintain a constant temperature measured by a PT100 sensors. The reduction of the power consumption of the resistor, in turn, enables the determination of the beam intensity at known beam energy.
\begin{figure}[htbp]
\centering
	\includegraphics[width=12cm,height=2cm]{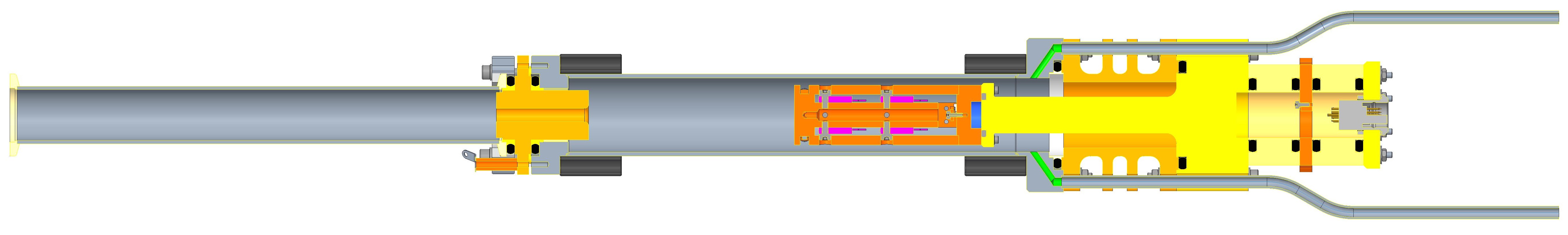} 
	\caption{Extended gas target chamber and a top view of beam calorimeter are shown. Two grey pipes at the right side of the figure, to inlet the gas inside the chamber.}
	\label{fig:3}
\end{figure}

\begin{figure}[htbp]
\centering
	\includegraphics[width=12cm,height=1.5cm]{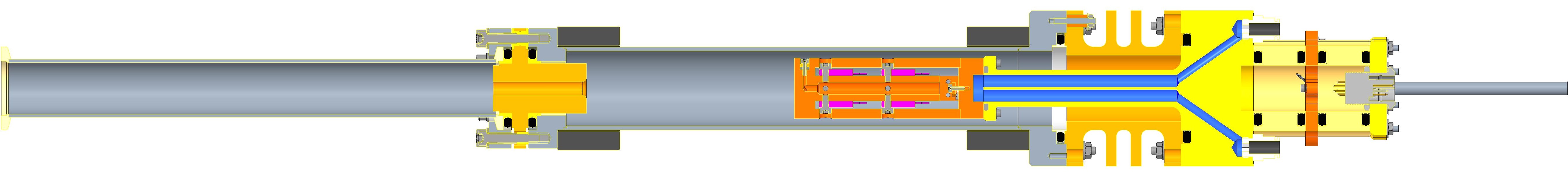} 
	\caption{Extended target chamber with the side view of beam calorimeter are shown. Two blue pipes at the right side of the figure are for water cooling of calorimeter.}
	\label{fig:4}
\end{figure}

\section{Measurement}
\subsection{Jet density measurement}
For a continuous gas jet density measurement, a laser interferometry technique \cite{3} is used, where the presence of gas induces an optical path length difference between the signal arm (passing through the jet) and the reference arm resulting in a phase shift on the interferogram. The optical path difference depends on the density distribution of the gas jet and its refractive index. The laser beam is expanded to illuminate the entire gas jet with the help of lenses to see the shape of the jet.

With the use of laser interferometry, the relative density and the shape of a jet are measured. The absolute density is determined by the energy loss method, where the energy loss of alpha particles traversing the jet are combined with tabulated experimental stopping powers at different points of the jet. For this absolute thickness measurement, a silicon detector will be mounted at the end of the beam line at one side of the jet and at the other side a triple alpha source (${^{239}}$Pu, ${^{241}}$Am, ${^{244}}$Cm) will be mounted. In this way, the energy of the alpha particles are measured with and without the jet and the difference in energy gives the energy loss of the alpha particles in the jet and hence the absolute density.
\begin{figure}[htbp]
\centering
	\includegraphics[width=10cm,height=5cm]{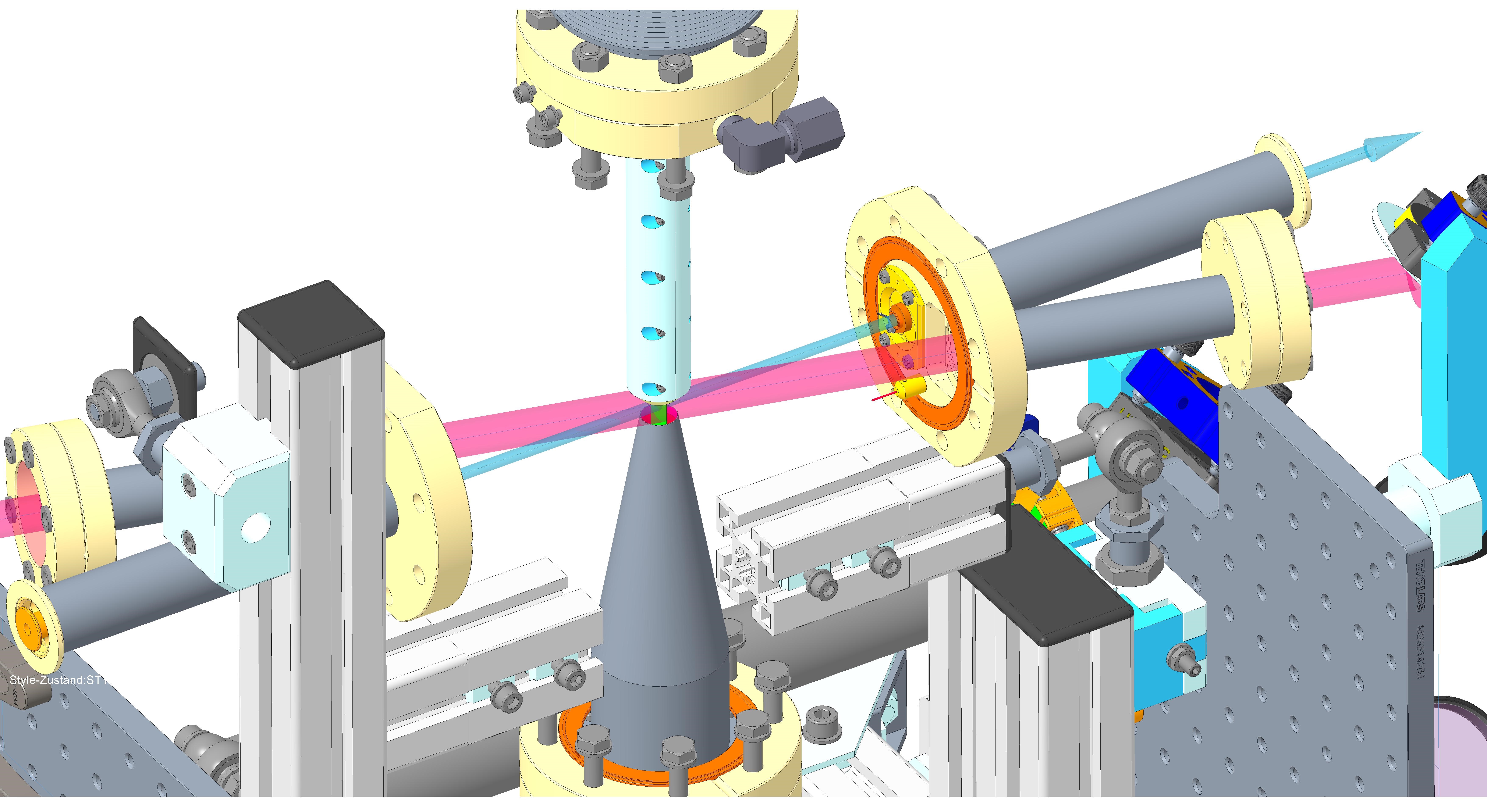} 
	\caption{Technical drawing of the gas jet interferometry setup. the laser and beam are showing in the red and blue colour respectively. The laser is passing through the jet at 15$^\circ$ with respect to the beam axis inside a jet chamber that is not shown here for the clarity.}
	\label{fig:5}
\end{figure}
\subsection{Pressure profile measurement}
\begin{figure}[htbp]
\centering
	\includegraphics[width=12cm,height=7cm]{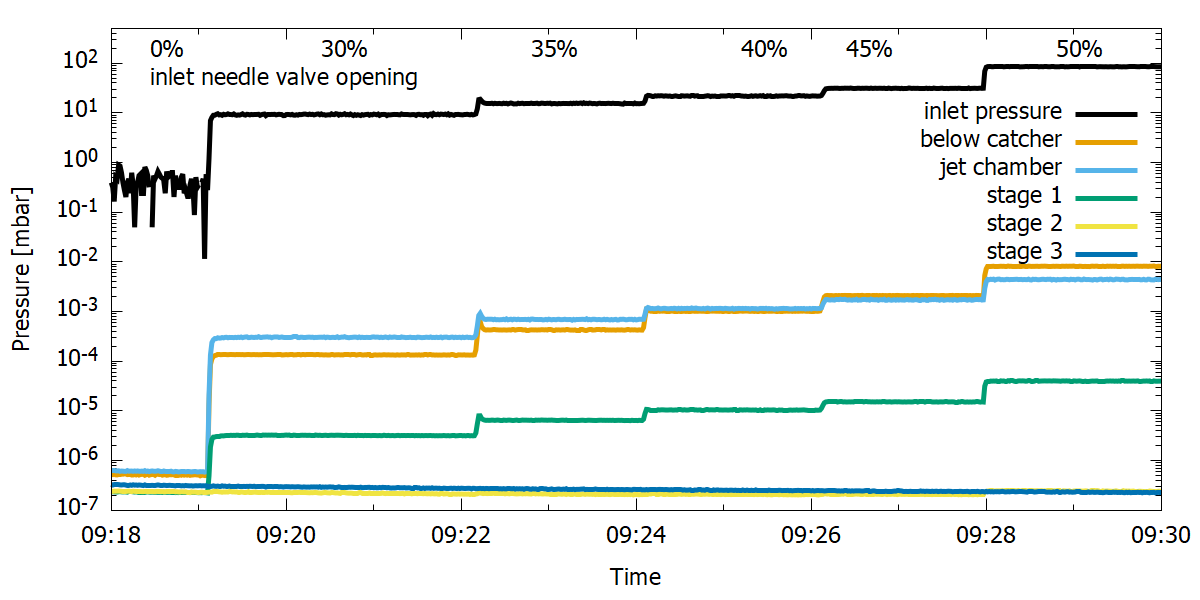} 
	\caption{Pressure profile measurements inside the chamber and beam line at different needle valve opening connected to the nozzle.}
		\label{fig:6}
\end{figure}
For first jet tests, a needle valves is placed at the nozzle inlet to vary the gas flow rate.  To monitor the pressure at different regions, pressure gauges are placed on all the pumping stages, the lower and upper cones of the expansion chamber, the nozzle inlet, and at the exhaust of all pumps. The pressure profile across the gas target setup at different needle valve opening for a first jet test with a preliminary nozzle are shown in Fig. 5. The pressure of the system without an inlet gas is in the order of $10^{-7}$ mbar in the beam line and expansion chamber. The pressures at the second and third pumping stages are not changing for different openings of needle valves. 

At an opening of 30$\%$, the pressure in the jet chamber is higher than below the catcher, indicating that there is more gas diffusing inside the chamber than going directly in the catcher. This is caused by the too small opening of the needle valve that is reducing the amount of gas to an inlet pressure of $\sim$10 mbar. This low inlet pressure is not able to accelerate the gas to supersonic velocity and hence not creating a jet. This behaviour changes at an opening of 50$\%$ when the inlet pressure increases to $\sim$100 mbar and most of the gas streams as a jet inside the gas catcher where about twice as much gas flows compared to the gas chamber. At an opening of 100$\%$ (not yet tested) the inlet pressure will reach 1 bar and create a supersonic jet into the catcher, with less then 1$\%$ of the gas drifting into the jet chamber. 

\section{Outlook}
The combined gas target has been built at the HZDR Rossendorf campus and will be transferred to the Felsenkeller underground laboratory in 2023. First jet tests have been performed to start a series of jet characterizations regarding pressure, density, and thickness. These are continuously measured by an interferometry setup that is in turn calibrated by an energy loss measurement of alpha particles across the jet.

There is further development of enhanced nozzles. Computational Fluid Dynamic (CFD) simulations in ANSYS Fluent are currently being carried out to optimize the shape of those upcoming nozzles.

Finally, in-beam tests are foreseen, as  soon as the target system is attached to the beam line of the 5-MV Pelletron accelerator. This combination of a wall-jet and a static gas target will be a unique in nuclear astrophysics and might become a role model for future gas target developments.

\section{Acknowledgement}
Financial support by the European Union (ChETEC-INFRA, 101008324) is gratefully acknowledged.

\end{document}